\documentclass[nonacm]{acmart}
\usepackage{array}
\usepackage{xcolor}
\AtBeginDocument{%
  \providecommand\BibTeX{{%
    \normalfont B\kern-0.5em{\scshape i\kern-0.25em b}\kern-0.8em\TeX}}}

\begin{document}

\title{Blockchain for Genomics: A Systematic Literature Review}

\author{Mohammed Alghazwi}
\email{m.a.alghazwi@rug.nl}
\orcid{0000-0003-4039-4748}
\affiliation{
  \institution{University of Groningen}
  \city{Groningen}
  \country{Netherlands}
}
\author{Fatih Turkmen}
\email{f.turkmen@rug.nl}
\orcid{0000-0002-6262-4869}
\affiliation{
  \institution{University of Groningen}
  \city{Groningen}
  \country{Netherlands}
}
\author{Joeri van der Velde}
\email{k.j.van.der.velde@umcg.nl}

\orcid{0000-0002-0934-8375}
\affiliation{
  \institution{University Medical Center Groningen}
  \city{Groningen}
  \country{Netherlands}
}
\author{Dimka Karastoyanova}
\email{d.karastoyanova@rug.nl}
\orcid{0000-0002-8827-2590}
\affiliation{
  \institution{University of Groningen}
  \city{Groningen}
  \country{Netherlands}
}

\renewcommand{\shortauthors}{Alghazwi et al.}
\renewcommand{\shorttitle}{Blockchain for Genomics: A Systematic Literature Review}

\begin{abstract}
Human genomic data carry unique information about an individual and offer unprecedented opportunities for healthcare. The clinical interpretations derived from large genomic datasets can greatly improve healthcare and pave the way for personalized medicine. Sharing genomic datasets, however, pose major challenges, as genomic data is different from traditional medical data, indirectly revealing information about descendants and relatives of the data owner and carrying valid information even after the owner passes away. Therefore, stringent data ownership and control measures are required when dealing with genomic data. In order to provide secure and accountable infrastructure, blockchain technologies offer a promising alternative to traditional distributed systems. Indeed, the research on blockchain-based infrastructures tailored to genomics is on the rise. However, there is a lack of a comprehensive literature review that summarizes the current state-of-the-art methods in the applications of blockchain in genomics. In this paper, we systematically look at the existing work both commercial and academic, and discuss the major opportunities and challenges. Our study is driven by five research questions that we aim to answer in our review. We also present our projections of future research directions which we hope the researchers interested in the area can benefit from. 
\end{abstract}


\maketitle

\section{Introduction}
The field of genomics holds great potential in enhancing healthcare. The genomic data produced through technologies such as high-throughput sequencing (HTS) can provide unique health-related information about each individual in a non-invasive manner and is crucial in the advancement of precision medicine~\cite{caskey2018precision}. It has been estimated that by 2025, between 100 million and as many as 2 billion human genomes could be sequenced ~\cite{stephens2015big}. At the same time as the amount of genomic data is soaring and genomic technology is advancing, so are the challenges they pose. Who has access to the data~\cite{schickhardt2020patients} and how can they be kept safe? How can data be used and shared responsibly \cite{phillips2018international} without losing the advantages of sharing for research and (future) patients? “The obligation to confidentiality” must be balanced with “the obligation to share” when it comes to genomic data. Some argue that sharing genomic data is an ethical obligation for those who benefited from sequencing their genome \cite{johnson2020rethinking}. However, genomic data have certain characteristics that make them fundamentally different from traditional health records: they are long-lived as they carry valid information even after an individual passes away; they indirectly affect descendants and relatives of the data owner, and they are large when the whole genome is sequenced (e.g. BAM file size is 138 GB~\cite{strandNGS}). 

While the potential societal impact of the improper use of genetic information is immense, there is a significant public benefit in the adoption of genomic data usage in patient diagnosis, screening, and treatment. The main technical issues that have been highlighted in the literature are the storage and sharing~\cite{ashley2016towards} of genomic data. More specifically, the need for efficient compression techniques, the lack of harmonized (meta)data, and perhaps more importantly the lack of secure and privacy-preserving technical infrastructures to acquire, process, and share genomic data. Another challenge is the lack of common terms and conditions in the metadata, which describes both the data (raw data format and notations) and the access requirements (informed consent and data transfer agreements). All these issues decrease the efficiency of discovery and hinder data sharing~\cite{wilkinson2016fair} in genomics research.

\subsection{Purpose and Objectives}
There is an increasing interest in applying blockchain technology in healthcare. This is evident in the increasing number of articles published each year since the emergence of this technology. For instance, \cite{holbl2018systematic} provided an overview of the current trends in using blockchain in healthcare and showed the properties of blockchain that are most commonly used, and \cite{agbo2019blockchain} classified each work in applying blockchain in healthcare based on the use-case. 
Blockchain has also been proposed as a candidate approach to address many of the challenges in handling genomic data~\cite{ozercan2018realizing, shabani2019blockchain, thiebes2020beyond}. The decentralization, immutability, and transparency properties make it an attractive option to solve the sharing and storage issues. In addition, it can be combined with privacy-preserving techniques to provide privacy, traceability, and integrity to the data being managed. 

In this review, we do not cover blockchain in general healthcare applications; our focus is on a subset of healthcare applications, namely, genomics. Our focus is on applications and solutions that utilize blockchain technology in managing, sharing, or processing human genomic data. Throughout the paper, we use the term "genomic data" to refer to data used in studying the human genome which includes the scientific study of gene interactions and complex traits/diseases that are caused by a combination of genetic and environmental factors. In addition, we use blockchain and DLT interchangeably even though there is a difference in the definition which is described in section \ref{block_background}. The use of blockchain in genomics is believed to have great potential as various researchers have pointed out. These works, as shown in Table \ref{tab-genomic-blockchain-articles} focused on the potential and possible benefits of blockchain in genomics, and they list the possible use-cases with a few example works that have been carried out. For instance, a recent paper~\cite{thiebes2020beyond} focused on exploring the opportunities and challenges of Distributed Ledger Technology (DLT) in genomics by conducting a ranking-type Delphi study. However, instead of focusing on the potential and possible use-cases, the main objective of this paper is to present the current state-of-the-art methods in applying blockchain in the field of genomics and the motivations for its use. In our study, we consider a wide range of commercial/non-commercial applications pertinent to genomic data and categorize the existing work. Finally, we discuss the limitations and future directions, which can serve as a basis for other researchers.

\begin{table}
\tiny
\caption{Reviews on the use of blockchain in genomics}
\label{tab-genomic-blockchain-articles}
\begin{tabular}{ccp{6.5cm}}
\toprule
\textbf{Paper} & \textbf{Focus} \\
\midrule
\cite{mackey2019fit, roman2018blockchain} & Potential of blockchain in healthcare  with genomics being discussed as an example. \\
\midrule
\cite{ozercan2018realizing} & Potential of blockchain in genomics and a use-case proposal. \\
\midrule
\cite{shabani2019blockchain, thiebes2019distributed, thiebes2020beyond} & Potential of blockchain in genomics.\\
\midrule
This paper & Present a comprehensive systematic literature review of the current state-of-the-art methods in applying blockchain in genomics. \\
\bottomrule
\end{tabular}
\end{table}

\subsection{Review Outline}
The review is structured as follows. First, the methodology used to conduct this review is explained in Section \ref{methodology}, then we give an overview of blockchain technology and genomic data storage/sharing in Section \ref{Background}. In Section \ref{results}, we summarize the findings of this review by describing the current trend, application domains, and motivations for using blockchain in genomics. In addition, we discuss the different approaches used and give an overview of the challenges faced. Section \ref{discussion} provides a discussion on our findings and the unexplored opportunities of applying blockchain in genomics.

\section{Review Methodology} \label{methodology}
\begin{figure}[h]
  \centering
  \includegraphics[width=8cm]{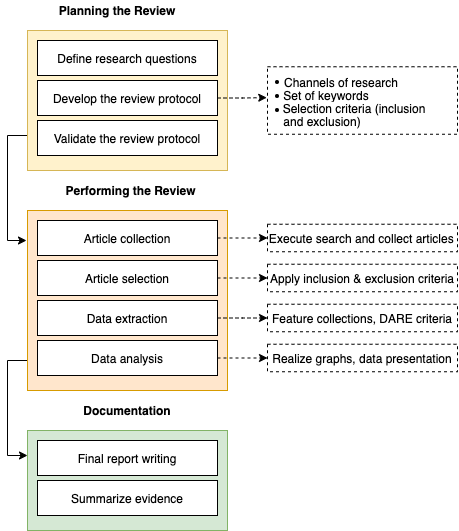}
  \caption{Methodology steps adapted from \cite{calvaresi2017exploring}.}
  \Description{}
  \label{fig:methods}
\end{figure}

The procedure used in this paper is in line with the methodology proposed by Kitchenham et al \cite{kitchenham2009systematic}. We utilize the Systematic Literature Review (SLR) approach as it provides a clear and structured way to search the literature and extract relevant information. Figure \ref{fig:methods} represents the methodology followed, which is adapted from \cite{calvaresi2017exploring}. The three main stages of the methodology are: planning the review, performing the review, and documentation. 

\subsection{Research Questions}
We formulated the main question "What are the application domains, motivations, approaches, and challenges when applying blockchain in genomic applications?" which can be split into the following specific and structured questions that will be addressed in this review: \\
\begin{itemize}
    \item RQ1: What are the current research trends for the use of blockchain in genomics?
    \item RQ2: What are the application scenarios of using blockchain in genomics?
    \item RQ3: What are the benefits and advantages of using blockchain in these applications as described by the authors? 
    \item RQ4: What are the elements of blockchain technology used in genomic applications? What are the approaches or combinations of technologies used?
    \item RQ5: What are the challenges and limitations when applying blockchain in genomics? Have these limitations been addressed by the authors? What has been specified for future research?
\end{itemize}

\subsection{Search Protocol}
The selected data sources include both academic and non-academic sources. This is done to cover a wide range of applications and approaches used in both academia and industry. For academic sources, we collected papers from the following 6 electronic databases: Google Scholar, IEEE Xplore, PubMed, Springer SpringerLink, Elsevier ScienceDirect, and ACM Digital Library. The search for relevant publications in these databases was performed using the following query strings:
\begin{verbatim}
(blockchain OR “block chain” OR "distributed ledger" OR "smart contracts") 
AND
(genomic OR genome OR genomics OR genes OR genetic OR genetics)
\end{verbatim}

In addition, preprints were collected from (arxiv.org). For non-academic sources, we used the Google search engine to find reports, blogs, and code repositories to select the relevant materials. This is done to find ongoing industry projects that considered the use of blockchain in genomics for commercial purposes. The set of keywords used to find these sources are selected based on the reviewers’ background and knowledge related to blockchain and genomic data sharing. These keywords include the following: \textit{genomics}, \textit{genomic data-sharing}, \textit{blockchain}, \textit{DLT}, \textit{smart-contracts}. The search was conducted in June 2022 and covered publications in the period 2009 - 2022.

\subsection{Selection Criteria}
The selection criteria (inclusion and exclusion) were defined prior to performing the search strategy in order to eliminate non-relevant sources. The selection criteria are shown in Table \ref{selection}.

\begin{table}
\tiny
\caption{Selection Criteria}
\label{selection}
\begin{tabular}{p{4cm}p{4cm}}
\toprule
\textbf{Inclusion Criteria} & \textbf{Exclusion Criteria} \\
\midrule
 1. Original research study (including grey literature). & 1. Secondary research, review papers, and non-relevant publications.\\
 2. Original work on the topic of blockchain in genomics. & 2. Publications presenting a point of view, magazine publications, interviews, and discussion papers. \\
 3. Publication years in the range between 2009 and 2021. & 3. Publications not in English. \\
\bottomrule
\end{tabular}
\end{table}

\subsection{Article Selection and Data Extraction}
\begin{figure}[h]
  \centering
  \includegraphics[width=10cm]{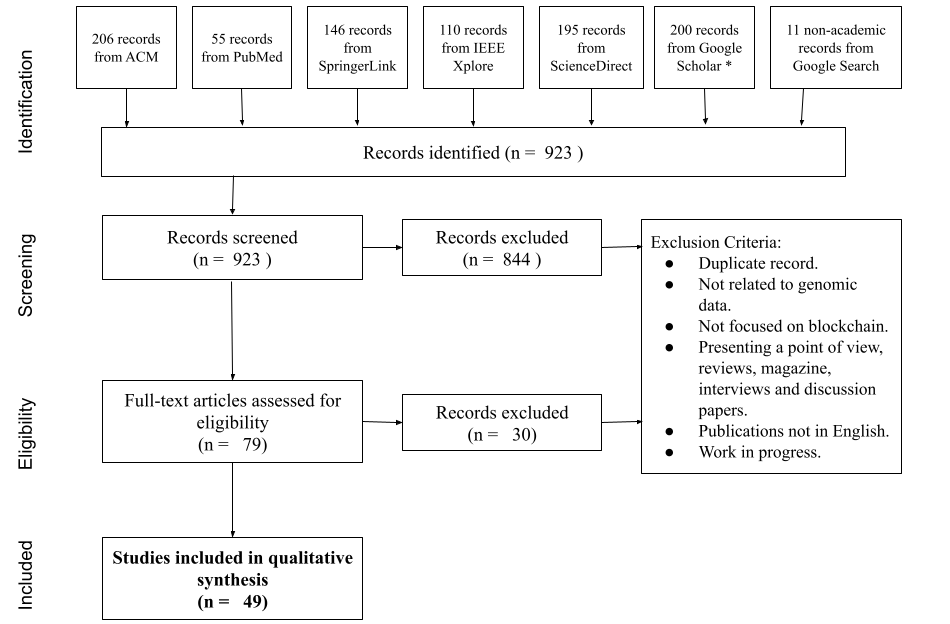}
  \caption{Identification and selection process}
  \Description{}
  \label{fig:identification}
\end{figure}

The initial search in the selected databases resulted in 923 papers. First, a screening of the titles and abstracts was performed with the aim to find and exclude duplicate and unrelated articles. The process reduced the number of relevant articles to 79. An additional analysis of the full-text was performed over these articles which resulted in 30 articles being discarded. The remaining 49 articles were included in this review as shown in Figure \ref{fig:identification}. We extracted data from each paper to determine the following: (1) the specific blockchain application that researchers focused on. (2) The motivations and advantages of using blockchain. (3) The approach/technology used to implement blockchain for genomic applications, including choices on the blockchain platform, storage, and security/privacy techniques. (4) The limitations and challenges regarding the design and implementation of the proposed application. 

\section{Background}
\label{Background}
In this section, we present background information on genomic data sharing and the key elements of blockchain. 

\subsection{Genomic Data: An Overview}
There are more than 5000 diseases for which a risk level can be calculated by using the genetic information of an individual according to DisGeNET~\cite{DisGeNet}. The genomic data presents an invaluable, unique source of information for understanding complex traits and diseases~\cite{NatureGenetics}. Traditionally, genomic sequencing (in particular whole genome sequencing) is considered to be a costly and time-consuming process. Thanks to the developments in genome sequencing technology, today, the time associated with whole-genome sequencing is at the level of hours (e.g. one sample in one hour~\cite{IMECInt}) and the cost is less than 600 Dollars~\cite{CNBC21}. Large datasets containing genomes and clinical data of individuals are becoming increasingly important to medical experts as the analysis of diverse data contributes to detecting fine-grain biological insights essential to improving public health~\cite{berger2019emerging}. As a result, an increasing number of clinicians are including analysis results obtained with these technologies in their day-to-day practices in the context of e.g. personalized medicine. Some of the frequent medical uses of genetic information include diagnostic and predictive DNA testing with the option of integrating polygenic risk scores where an individual’s (and her relatives’) disposition to certain diseases such as breast cancer is screened through specific genes (e.g. BRCA2).

\subsubsection{Genomic data sharing}
Interpretation is a key component of genomics research. 
Individual genomic variants can be interpreted in relation to specific signs or symptoms, and multiple genomic variants can be assessed in relation to their collective impact on patients. Genetic (genotype) and clinical data (phenotype) can be combined to determine the best treatment for a patient. The quality of these interpretations is highly dependent on the data they are based on. Research and clinical knowledge sharing are essential to enable the refinement of interpretations.

While policies and laws, which differ from country to country, allow the exchange of genomic data under certain conditions, genomics researchers have experienced how difficult and cumbersome the process is \cite{learned2019barriers}. Another issue is the participation in genomics research which is currently low considering the requirements in genomics research. For instance, the number of participants (i.e., subjects) in genome-wide association studies (GWAS) can reach over 1 million \cite{mills2019scientometric}. 

There are various genomic data exchange platforms that give researchers the ability to share genomic data publicly in order to advance research in this field. Large organizations such as Clinical Genome (ClinGen)~\cite{clingen} and the Global Alliance for Genomics and Health (GA4GH)~\cite{ga4gh} have started the development of reliable resources to systematically define and interpret all human genome variations through broad data-sharing efforts. There are also large-scale European efforts to promote/coordinate the cross-border collection, storage, and sharing of human genome data in a secure way, e.g. Beyond 1 Million Genome (B1MG)~\cite{Beyond1MG}. One of the most prominent solutions to large-scale genomic data sharing is the Genomic beacon project initiated by GA4GH \cite{fiume2019federated}. Genomic beacons ease the process of data sharing through the use of web services by answering queries about the presence of a specific allele in a genome. Institutions can launch their own beacons and connect to the project. In general, beacons aim to respect data privacy by allowing the institutions to define their own access restrictions and authorization schemes. However recently, researchers showed that individuals can be identified even if a data anonymization technique is applied to the data~\cite{shringarpure2015privacy} that is made available through beacons. 

\subsubsection{Privacy of genomic data} 
In addition to being highly valuable, genomic data is highly sensitive as it may reveal information about an individual and/or his/her family. Current genomic data sharing methods are highly dependent on the level of privacy (and the task at hand) required as some parts of the genomic data are private while some others are not considered private. For instance, somatic variants in the human genome are not considered private in some contexts as they cannot identify specific individuals or families. On the other hand, germ-line variants are unique to each person, and therefore, they require privacy protection in many use cases. The susceptibility to certain diseases, ancestral traits of an individual, and response to a drug are just a few of the use cases demonstrating the private nature of one's genetic information.

The conflicting need for genomic data to be both shared and private requires the use of privacy-preserving techniques that allow processing of the data without the identification of individuals. There is a plethora of research on the topic of privacy{-}preserving processing of genomic data that propose the use of privacy-enhancing technologies such as Homomorphic Encryption~\cite{Dowlin2017, Ayday2013} and secret sharing~\cite{Zhang2015,Sotiraki2020} among others. We refer the reader to the surveys on the topic such as~\cite{akgun2015} for a more systematic presentation.

The large size of genomic data, which ranges between 30-200GB, is one of the main obstacles in the application of privacy-preserving techniques to genomic data processing. There are proposals to substantially reduce the storage footprint of genomic data. For instance, PetaGene~\cite{Petagene} proposes a compression technique on NGS datasets in FASTQ and BAM formats and aims to provide on average 60\% reduction without genotyping accuracy loss. However, with the large-scale collection efforts at the horizon such as B1MG~\cite{Beyond1MG}, an explosion of genomic data is expected. This opens the field to alternative approaches that comply with the laws and regulations around privacy and at the same time preserve the utility of the data. \\

\subsection{Overview of Blockchain}
\label{block_background}
The detailed technical foundation of blockchain technology is outside the scope of this paper. However, it is important to shed light on some blockchain concepts, features, and terminologies that will assist the understanding of how blockchain is applied to solve problems in handling genomic data. For an extensive treatment, we refer the reader to other articles such as Kolb et al. \cite{kolb2020core}. \\

Although blockchain technology and Distributed Ledger Technology (DLT) are closely related, there is a difference. A distributed ledger is a ledger or a database that is spread across the nodes in the network and maintained by a group of peers, rather than a central agency. Blockchain is an implementation of DLT and unlike a database, it consists of a chain of blocks. These data blocks are unique data structures that distinguish blockchains from other DLT types. Other implementations of DLT include Hashgraph \cite{baird2016swirlds} and Directed Acyclic Graph (DAG) \cite{vzivi2019directed}. Blockchain is the innovative technology behind Bitcoin~\cite{nakamoto2008peer}, the first open-source decentralized digital currency system. Blockchain stores and verifies transactions on a ledger that is distributed to all nodes in a peer-to-peer (P2P) network. The transactions are organized into blocks that are protected by a combination of cryptographic techniques to ensure the integrity of the recorded transactions. A consensus protocol is then followed to validate the blocks and the blocks that are successfully validated are added to the growing chain of blocks. This essentially solves the problem of allowing multiple parties that do not necessarily trust each other to agree on the state of a shared ledger. Proof-of-work (PoW) in bitcoin was the first consensus protocol used in blockchain. PoW has been criticized for its energy waste and slow block confirmation. Various protocols have been developed to overcome some of the limitations in PoW such as Practical Byzantine Fault Tolerance (PBFT) and Proof-of-stake (PoS) \cite{xiao2020survey}.

\subsubsection{Blockchain types}
There are three distinct types of blockchains and they differ in how the nodes can join the network and how the network behaves. These types of blockchain can be classified into the following. We refer the reader to \cite{sultan2018conceptualizing} for further discussion on this classification. 

\begin{enumerate}
  \item Public Blockchain (public \& permissionless): a public blockchain is open to anyone (and thus trustless). Any anonymous node can join the network, and no trust requirement is enforced by the network members. The transactions are publicly broadcasted to all the nodes. Any node in the network can participate in the consensus mechanisms to validate the blocks. An example of this type of network is the Bitcoin blockchain \cite{nakamoto2008peer}. 
  \item Private Blockchain (private \& permissioned): nodes are pre-selected and vetted, and once they join the network they are trusted. The access control mechanism provides a higher degree of privacy to the content of the blockchain transactions. Additionally, this type of blockchain provides higher performance in terms of block confirmations. An example of a platform of this type is MultiChain \cite{multichain}.
  \item Consortium Blockchain (public \& permissioned): a mix of the previous two by offering a compromise between efficiency and trust. The ledger that is managed by a pre-selected group of nodes. Similar to private blockchains, nodes require permission to join the network. Validating blocks and transactions is done when a chosen set of nodes reach a consensus. The exact process depends on the pre-established rules of the consensus mechanism. An example of a platform of this type is Hyperledger \cite{cachin2016architecture}.
\end{enumerate}

\subsubsection{Smart Contracts} 
Smart contracts have emerged recently on blockchain due to the popularity of the Ethereum platform. However, the concept of smart contracts dates back to 1997 and was proposed by Nick Szabo~\cite{szabo1997formalizing}. The concept has evolved since then, but the main objective is to allow a smart contract program to run in a decentralized network and modify the state of the system in an automated, trusted, and verifiable way without intermediaries. Blockchain has made it possible to implement this concept and use it in different settings including finance, Identity Management, and healthcare \cite{khan2021blockchain}. There is a variety of blockchain platforms that allow executing smart contracts using a number of programming languages and one of the most commonly used languages is Solidity. Briefly, writing a smart contract involves establishing a set of requirements and instructions that are automatically executed once these requirements are met. In contrast to written contracts, smart contracts are executed automatically, are publicly verifiable, and do not require any intermediaries.

\subsubsection{Security and Privacy in Blockchain}
The security and privacy features of blockchain rely on the use of a number of cryptographic techniques. Some of these techniques were leveraged by the original Bitcoin blockchain design, while others were added to subsequent blockchain implementations to enhance security and privacy. The basic security and privacy techniques utilized in the bitcoin blockchain ensure that the system meets the security and privacy related requirements of online transactions. Blockchain ensures the consistency of the ledger across multiple nodes through the use of a consensus mechanism discussed previously. Blockchain transactions are resistant to tampering from both miners that confirm those transactions, and the external attackers that try to manipulate blockchain transactions. Using cryptographic hashing and digital signature, any modification on the transaction data would be detected by checking the validity of the digital signature. In addition, tampering with blockchain transactions requires altering the data stored in all blockchain nodes since the ledger of transactions is stored in all nodes in the network. Attacks such as distributed-denial-of-service (DDoS) are not feasible because of the highly decentralized nature of the blockchain network.

Regarding the privacy aspect, blockchain provides pseudonymity through the use of public key infrastructure (PKI). The nodes or users are identified by their public addresses rather than their real identities. However, this fails to provide full anonymity as there is a risk of linking the public address to real identities by observing the interactions between different parties \cite{zhang2019security}. Another privacy limitation of the original blockchain implementation is that transactions and their data are publicly visible. Ensuring the confidentiality of transactions and smart contract data expands the possible applications of blockchain to include those that handle private and sensitive information. To overcome the aforementioned limitations, additional security and privacy techniques have been proposed such as mixing, anonymous signatures, and Zero-Knowledge Proofs (ZKP) \cite{zhang2019security}.

\section{Results} \label{results}
In this section, we present the outcomes of the review. The aim is to answer the research questions defined in Section \ref{methodology}. We first show the distribution of publications per year. Then, we list and classify each paper into sub-categories based on the application domain, and present our analysis on the motivation of using blockchain as described in the papers. We also present the methods and approaches employed in these papers. Finally, we list the open issues and challenges we identified through our study. 

\subsection{Current Trend in Genomic Blockchain Research}

\begin{figure}[h]
  \centering
  \includegraphics[width=10cm]{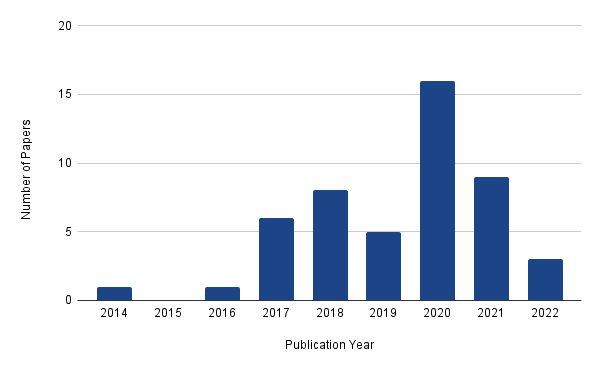}
  \caption{Number of publications per year presenting a proposal on the use of blockchain in genomics. The bars show the number of papers for each year}
  \Description{}
  \label{fig:trend}
\end{figure}

In this section we aim to answer RQ1: What are the current research trends for the use of blockchain in genomics? We analyzed the yearly trend in publications related to the use of blockchain in genomics. As can be seen in Figure \ref{fig:trend} there is an increasing interest in this topic. The interest started with a lot of commercial applications, but with time more academic research followed. Based on our findings, the first use of blockchain in genomics appeared in a commercial application, Genecoin~\cite{genecoin}, which started in 2014. The number of papers increased over the years and this quick upward trajectory seen in Figure \ref{fig:trend} is expected since blockchain is a relatively new technology that was introduced in 2009 and the implications of its use (e.g. for scalability and security/privacy) are just being studied in a non-cryptocurrency context. 

\subsection{Blockchain in Genomic Applications}
\begin{figure}[h]
  \centering
  \includegraphics[width=10cm]{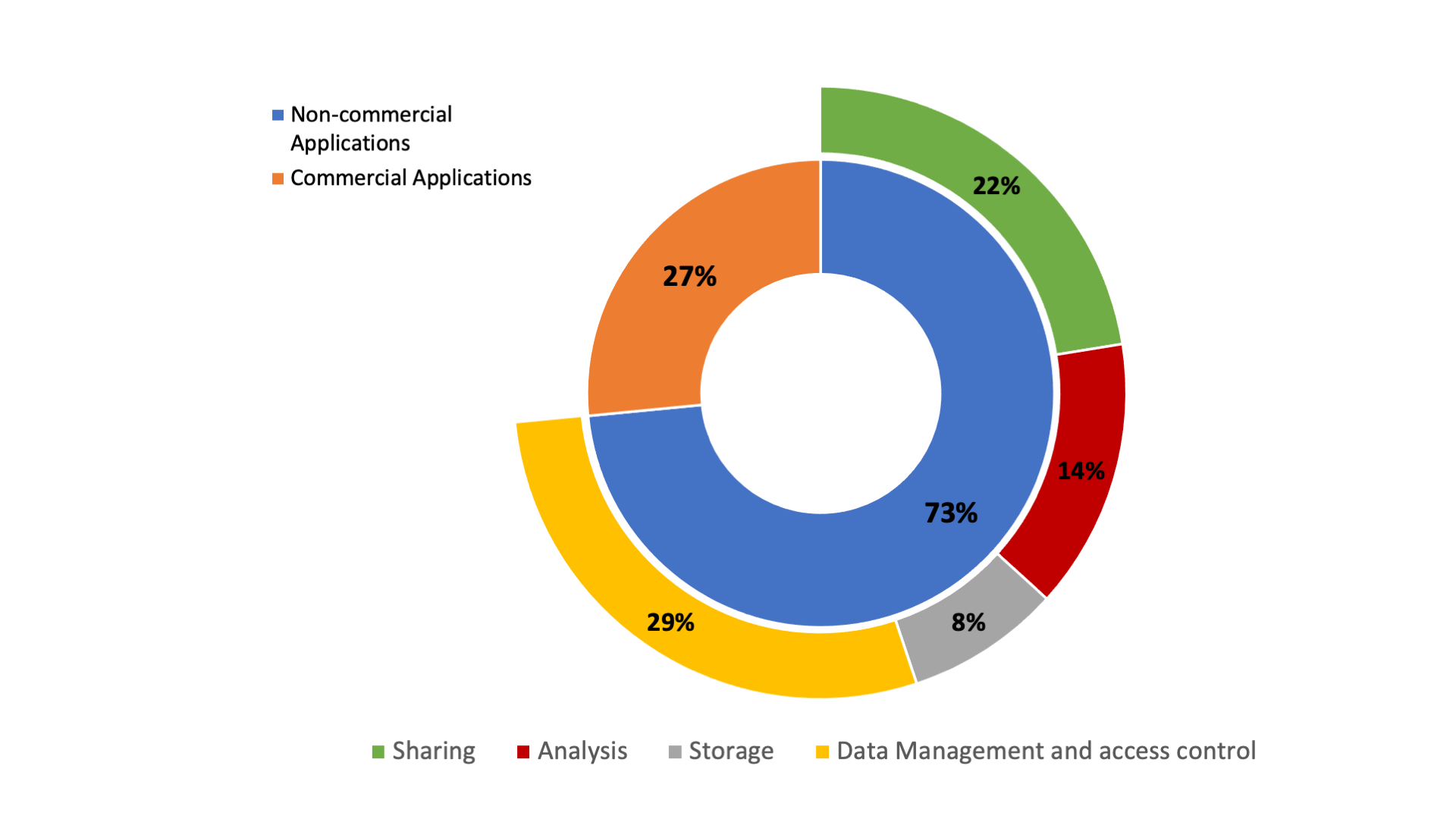}
  \caption{Classification of genomic blockchain applications}
  \label{fig:cetegories}
\end{figure}

This section reports the range of existing blockchain-based solutions in genomics, which answers RQ2:What are the application scenarios of using blockchain in genomics? Because the use of blockchain in genomics has attracted the attention of both academic and industrial communities, each with their agenda on how this technology can be used, we classified the applications as shown in Figure \ref{fig:cetegories} in two main categories, namely, \textit{commercial} and \textit{non-commercial}. To clarify our classification process further, we distinguish between the two categories based on whether the blockchain is utilized for financial exchange in addition to data sharing. Commercial genomic marketplaces follow a business model and are generally aimed at facilitating the exchange of genomic data for financial benefits. In addition, these marketplaces are usually targeted at individual genomic data owners (individual users/customers or patients), and cryptocurrency or tokens are often used as incentives to promote data sharing. From the selected papers in our study (49), there are 36 papers with no commercial interests and 13 papers with a commercial motivation.

\subsubsection{Commercial Genomic Marketplaces}
The commercialization of DNA sequencing by direct-to-consumer (DTC) companies has attracted an increasing number of customers in recent years. This is due to the technological advances that made genome sequencing much cheaper and faster. One of the strategies to generate revenue for DTC companies such as 23andme is selling access to the collected DNA sequences to pharmaceutical companies. The fairness of this model raises questions in terms of the profit gained from buying this genomic data. Some argue that the profit should be passed onto the people, not the intermediaries \cite{shabani2019blockchain}. As a result, there has been an increase in a new generation of companies that provide an open marketplace for genomic data sharing with the use of blockchain.

Blockchain-based genomic marketplaces aim to cut the need for intermediaries and give users control of their data. Individuals receive different types of incentives for selling or renting their genomic data. The most common incentive used in these marketplaces is cryptocurrency. Table~\ref{tab-marketplaces} provides an overview of the used incentives along with the employed blockchain platform and the services offered by this platform.

\begin{table}
\tiny
\caption{Blockchain-based Commercial Genomic Marketplaces}
\label{tab-marketplaces}
\begin{tabular}{p{2.5cm}p{3cm}p{3cm}p{3.5cm}}
\toprule
\textbf{Company} & \textbf{Blockchain Platform} & \textbf{Incentive} & \textbf{Services} \\
\midrule
Genecoin \cite{genecoin} & Bitcoin blockchain & N/A & Encrypted DNA storage/backup in the Bitcoin blockchain. \\
\midrule
Encrypgen \cite{encrypgen} & GeneChain to facilitate cryptocurrency exchange and record data exchange & Cryptocurrency: DNA token & HIPAA-compliant storage of genomic data and exchange platform after de-identifying the data. \\
\midrule
Embleema \cite{embleema} & Private Ethereum blockchain to register patient consent and manage data flow & Cryptocurrency: RWE token & Patient consent management, optimized clinical trial recruitment, secure HIPAA-compliant storage, analysis and organization of the data for researchers. \\
\midrule
Nebula Genomics \cite{Nebulagenomics}, \cite{grishin2018accelerating}, \cite{grishin2019citizen} & Exonum \cite{Exonum} permissioned blockchain for decentralized access control logs, data storage in third-party storage system & Nebula credits which can be used to purchase health-related information about the genome sequence & User controlled access to genomic data with blockchain log, secure storage and analysis in a blackbox environment. \\
\midrule
LunaDNA \cite{lunadna} & Unspecified & Company shares and earnings when data are used & HIPAA-compliant storage, user ownership and control of data, contributions to research and medical research. \\
\midrule
Shivom \cite{shivom} & Blockchain agnostic based on combination of blockchain implementations & Cryptocurrency: OMX tokens & Pay-per-use marketplace for genomic data, bioinformatics platform with analysis pipelines for researchers. \\
\midrule
Zenome \cite{zenome} & Ethereum-based blockchain & Cryptocurrency: ZNA token & Secure storage, selling access to genetic data, and buying genetic services.\\
\midrule
Genomes.io \cite{genomes.io} & Ethereum-based Blockchain & Cryptocurrency: GENE token & Secure personal DNA vault, and financial benefits from (anonymous) contributions to medical research. \\
\midrule
DNAtix \cite{dnatix} & Ethereum-based blockchain (proof-of-concept) & Cryptocurrency: internal ERC-20 based token & Anonymous genetic vault service, anonymous testing and reports on genomic data, and the ability to connect anonymously with people with similar genetic traits. \\
\midrule
Genesy \cite{carlini2020genesy} & Private blockchain based on HyperLedger Fabric & Payment in fiat and cryptocurrency through Stellar \cite{stellar} and Stripe \cite{stripe} & Sequencing services, selling access to genomic data, and a blockchain-based ecosystem for the sharing of genomic data. \\
\midrule
GenoBank \cite{uribe2020privacy} & Ethereum-based blockchain with non-fungible tokens (NFT) & Cryptocurrency: ERC-20 token & Control over genomic data with DNA crypto wallet, and secure platform to process the data. \\
\midrule
Longenesis \cite{longenesis} & Bitfury’s Exonum blockchain & N/A  & Patient consent platform targeted to medical institutions; HIPAA and GDPR compliant data storage.  \\
\midrule
LifeCode.Ai \cite{jin2019application} & Quorum platform \cite{quorum} & Cryptocurrency: ERC-20 token & Individual health data ownership, trading mechanism for data exchange, and secure decentralized storage and management of health data\\
\bottomrule
\end{tabular}
\end{table}

Based on our findings we found that Genecoin \cite{genecoin} marked the first attempt at using blockchain in genomics when it was introduced in 2014. The company provides sequencing services through third-party labs. It then encrypts and stores the resulting DNA sequence in the bitcoin blockchain \cite{genecoin}. Many other marketplaces followed Genecoin with the goal of allowing genomic data owners to exchange their data for cryptocurrencies (tokens), which later can be traded in cryptocurrency exchanges. 

Most marketplaces in Table~\ref{tab-marketplaces} follow the model Genecoin put forward and with that they aim to empower patients by letting them control their data, and accelerate genomic research by enabling the availability of more data. Other marketplaces differentiate themselves by providing additional services, for instance, Shivom \cite{shivom} and Zenome \cite{zenome} provide pipelines and computational resources to perform analysis on the genomic data that is exchanged on their platform. Researchers can conduct their research through the provided pipelines by paying fees in cryptocurrency/tokens. In addition, data integrity and traceability are highlighted as essential services provided by some of the genomic marketplaces. For instance, Genomes.io \cite{genomes.io} allows consumers to securely store and manage their DNA data from the moment it is sequenced to when it is stored on the blockchain. The users are provided with a private access key that is stored only on their mobile devices and is used to control access to the data. Another example is Genobank \cite{uribe2020privacy} which explores the use of non-fungible tokens (NFT) for portability and data tracing. The proposed method is to assign each human sample (e.g. saliva) a unique NFT and share it in the public blockchain. The NFT is used for a variety of purposes including the decision on which lab can sequence the genome and the tracking of the use of the resulting data.

\subsubsection{Non-Commercial Applications}
The selected studies on non-commercial applications of blockchain mainly focus on providing genomic data sharing for the advancement of research. These studies fall into one of the classes of use cases in using blockchain: data sharing, data analysis, storage, and access control. While each paper is categorized according to the main topic of research, overlaps can occur. For example, one study focused on genomic data sharing for the purpose of performing analysis tasks on that data, and another paper focused on blockchain-based storage for the purpose of sharing the stored genomic data. In Table \ref{tab-taxo} we provide a taxonomy of non-commercial blockchain applications in genomics. The taxonomy aims at categorizing and summarizing the ideas and use cases of the selected papers in this review. In what follows, we discuss each scenario in more details with examples.

\begin{table}
  \tiny
  \caption{Taxonomy of genomic blockchain applications}
  \label{tab-taxo}
  \begin{tabular}{ccc}
  \toprule
  \textbf{Main Category} & \textbf{Focus} & \textbf{Papers/References} \\
  \midrule
  Data Sharing & Secure genomic data sharing platform & \cite{zhangy2018genie, dambrot2018regene, shuaib2020layered, iyer2019ai, mathur2020immutable, neto2020research, miyachi2021hocbs} \\
  \cmidrule(r){2-3}
  & Sharing and providing open-access to deidentified clinical and genomic data & \cite{Glicksberg2020, ozercan2018realizing} \\
  \cmidrule(r){2-3}
  & Consensus protocol for sharing health and genomic data & \cite{talukder2018proof, ileri2016coinami} \\
  
  \midrule
  Data Analysis & Decentralized privacy-preserving GWAS & \cite{zhang2019enabling} \\
  \cmidrule(r){2-3}
  & Decentralized privacy-preserving predictive model learning & \cite{kuo2019fair, kuo2020anatomy, kuo2020ex, kuo2020privacy, kuo2022detecting} \\
  \cmidrule(r){2-3}
  & Decentralized training of machine learning models on clinical and genomic data & \cite{warnat2021swarm} \\
  
  \midrule
  Storage & Storing raw genomic data on-chain & \cite{gursoy2020storing} \\
  \cmidrule(r){2-3}
  & Lossless compression for efficient storage and transmission of DNA sequence data & \cite{lee2018baqalc} \\
  \cmidrule(r){2-3}
  & Storage and query of pharmacogenomics data & \cite{gursoy2020usingethereum, kuo2021benchmarking} \\
  
  \midrule
  Data Management and access control & Access and consent management of personal genomic data & \cite{pachaury2021securing, kim2021blockchain, park2021blockchain, arroyo2021technological, ma2022phdmf} \\
  \cmidrule(r){2-3}
  & Multi-Stakeholder consent management & \cite{beyene2019multi} \\
  \cmidrule(r){2-3}
  & genomic dataset activity and access logs & \cite{gursoy2020usingblockchain, ma2020efficient, ozdayi2020leveraging, pattengale2020decentralized, sulistyawan2021data} \\
  \cmidrule(r){2-3}
  & Dynamic consent management & \cite{mamo2019dwarna, albalwy2021blockchain, silva2022idealized} \\
  
  \bottomrule
  \end{tabular}
  \end{table}

\textit{A. Data Sharing } \\
The majority of the papers we identified focused on using blockchain to support/build systems for multi-organizational or global sharing of genomic data. A few of these papers proposed different system architectures and platforms for sharing genomic data utilizing different aspects and benefits of blockchain \cite{zhangy2018genie, dambrot2018regene, shuaib2020layered, iyer2019ai, mathur2020immutable, neto2020research, miyachi2021hocbs}. Another subcategory of papers focused on the use of blockchain and its decentralized nature to support open-access and sharing of deidentified clinical and genomic data. A noteworthy work is the Cancer Gene Trust (CGT)~\cite{Glicksberg2020}, which demonstrates the benefits of using blockchain in sharing genomic data, for the purpose of advancing cancer research. In addition, the authors launched a cohort study with a real patient dataset to illustrate the effectiveness of the CGT framework in terms of secure, efficient, cost-effective, open, and distributed sharing of genomic data. The data for the pilot study was acquired from cancer patients with their consent, and de-identification was applied in order to preserve the privacy of the patients. The de-identification process was done because the data is intended to be publicly available for any other researchers to analyze. CrypDist \cite{ozercan2018realizing} presented a similar approach, but with additional mechanisms to distribute the whole-genome data. The last subcategory of papers focused on alternative consensus protocols that are specific for health and genomics applications. For instance, Coinami \cite{ileri2016coinami} provided an alternative approach by incentivizing participants to perform HTS read mapping on a voluntary basis. The participants are given tokens as a reward. This replaces the traditional proof-of-work with HTS read mapping in the validation of blocks before they are added to the chain.

\textit{B. Data Processing/Analysis} \\
A set of papers explored the use of blockchain in facilitating genomic data processing or analysis. Zhang et al \cite{zhang2019enabling} proposed an approach to perform a GWAS with an emphasis on privacy. In their approach, GWAS is performed by using a privacy-preserving sharing protocol that has a gene fragmentation framework at its core. In this framework, the large genomic files are split into multiple fragments which are then distributed to multiple service providers, that form a decentralized blockchain network, for storage, sharing, and analysis. This eliminates the possibility of one provider having the complete data, and therefore, solves the issues related to centralization and privacy. Other papers proposed combining machine learning with blockchain to achieve decentralized machine learning. In \cite{kuo2019fair, kuo2020anatomy, kuo2020ex, kuo2020privacy, kuo2022detecting}, predictive models were trained in multiple organizations with blockchain coordinating the process in a secure, transparent, and decentralized way. Additionally, federated learning using a blockchain solution called Swarm Learning (SL) was proposed in \cite{warnat2021swarm}. SL performs decentralized machine learning by first securely onboarding members without the need for a central coordinator. Then, the data are trained locally and the model parameters are merged thus keeping the large medical and genetic data in the possession and control of the data owner.

\textit{C. Secure storage} \\
Among the selected papers, we found a group of papers that focused on using blockchain as a way to store genomic data securely. \cite{gursoy2020usingethereum, kuo2021benchmarking} utilized blockchain to store and query pharmacogenomics data. The authors illustrated the feasibility of storing and accessing genomic data using the Ethereum blockchain and smart contracts. Each data record is inserted into the smart contract and is assigned a unique ID that is used as a mapping key. An index-based, multi-mapping approach is used to efficiently query the genomic data. The pharmacogenomics data used in this study is rather small in size compared to other common types of genomic data types. In \cite{gursoy2020storing}, the authors explored alternatives for storing larger data files, specifically Sequence Alignment Map (SAM) files, which can be in the order of 10s of Gigabytes in size. This was achieved with a novel data structure that was built with the combination of data compression techniques and a private blockchain network. 

\textit{D. Data Management and access control} \\
Another set of papers focused on using blockchain as a means to grant/revoke access to genomic data, which may be stored somewhere else, in the form of consent management. Dwarna~\cite{mamo2019dwarna} is a web portal that harnesses blockchain for dynamic consent. The portal connects participants and researchers in a research partnership. The project incorporates GDPR and aims to ensure that the ownership of the data remains with the participants. The proposed architecture uses blockchain to record participants' consent. Storing consent in blockchain allows the participants to be the owner of the data, i.e., third parties can only access the data when the owner of the data allows them to. In \cite{beyene2019multi}, the authors consider consent for sharing individual genomic data as an instance of the Multi-Stakeholder Consent Management (MSCM) problem. This is due to the fact that each individual genome can reveal information not only about the owner of that genomic data but also about the relatives. Therefore, to protect the privacy of the relatives of an individual, their consent must be taken into account. The authors in \cite{beyene2019multi}, propose the use of blockchain to solve this consensus problem and obtain consent from multiple stakeholders. The 2018 IDASH competition \cite{kuo2020idash} and specifically, Task 1 of the competition explored the use of blockchain as a global logging system. Such a system can be used to provide an access log that records users’ access to any data within any of the genomic data repositories in the system. A decentralized cross-site logging system has many advantages over traditional centralized internal logs that are currently common in practice. Most importantly, it eliminates the problem of a single point of failure and malicious changes to the logs. There were several participants \cite{gursoy2020usingblockchain, ma2020efficient, pattengale2020decentralized, ozdayi2020leveraging} in the competition and the submissions were evaluated based on specified criteria which include accuracy and speed. This is because the competition not only looked at the feasibility of blockchain as a cross-site logging system but also evaluated its performance and efficiency. The competition showed that it is indeed feasible to utilize blockchain in building a cross-site genomic data access log. The performance of that solution is promising, and it is reasonable to assume that with additional improvements, such a system can be adopted for practical use. 

\subsection{Motivations for the use of blockchain} 
\begin{figure}[h]
  \centering
  \includegraphics[width=10cm]{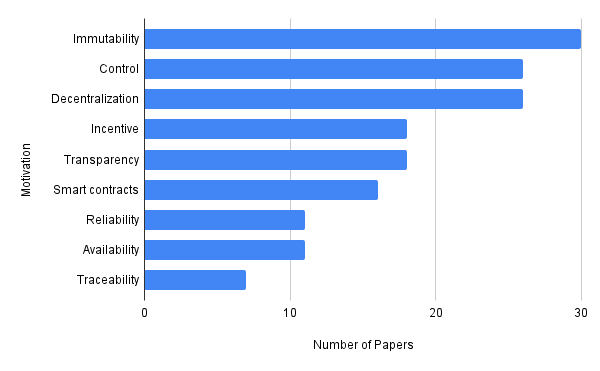}
  \caption{Frequency of motivations for the use of blockchain in genomics.}
  \Description{}
  \label{fig:motivations}
\end{figure}

To address RQ3: What are the benefits and advantages of using blockchain in these applications as described by the authors? we first identified the key blockchain features that are most desirable in genomic applications. These key features are incentives, decentralization, stringent control of data, immutability, smart contracts, reliability, availability, transparency, and traceability. The motivation depends on the outlined requirements that are listed in each paper and thus may be different for each genomic application. Figure \ref{fig:motivations} shows the frequency of each key blockchain feature that motivated its use in the selected papers. Most papers list multiple benefits of using blockchain and we account for this in Figure \ref{fig:motivations} by adding the papers to the count for each mentioned blockchain feature. For instance, one of the papers listed immutability and decentralization so this paper will be included in the count for both of these blockchain features. In general, the most highlighted feature is that blockchain is an immutable and tamper-proof way to store genomic data. In addition, decentralization and control of the data are highly mentioned benefits. The rest of this section provides a summary of the motivations to apply blockchain in the selected studies as described by the authors.

\textit{A. Incentive (Cryptocurrency).} An important aspect of using blockchain is the ability to build an incentive structure through cryptocurrencies or tokens for sharing genomic data. This is especially relevant for genomic marketplaces where the objective is to create a fair ecosystem for the exchange of private data. This fairness is defined in terms of financial gain from data sharing (by data owners), and the aim is to allow the exchange of data for scientific research or other purposes without losing full control. Additionally, an incentive scheme can be used to reward nodes in that network after completing a certain task such as sequence (HTS) read mapping in \cite{ileri2016coinami}. Any individual or organization is able to freely join and perform analysis tasks to gain tokens that could be redeemed for monetary value. 

\textit{B. Decentralization.} The decentralized nature of blockchain networks was listed as an important feature in several papers. The consensus mechanism contributes greatly to the way blockchain is decentralized. It introduces a way for nodes in the network to reach an agreement without a central trusted authority. Decentralization provides several benefits depending on the use case. In \cite{Glicksberg2020} decentralized open access is achieved by using blockchain combined with IPFS. The timely distribution of medical resources plays a significant role in the research and development of medical treatment, especially in the event of a disease outbreak such as COVID19. Decentralization is also useful in other cases such as coordination of analysis tasks that are performed at multiple nodes/locations. As shown in \cite{kuo2020privacy, kuo2020ex, kuo2020anatomy, kuo2019fair}, blockchain can replace the need for a central server to intermediate the process of applying machine learning and combining the global model. This prevents the single point of failure/control when a third-party is used to coordinate, and the potential for this third-party to breach the privacy of data by examining the aggregated statistics. A similar approach is shown in \cite{zhang2019enabling} where the motivation for using blockchain is to coordinate the process of performing GWAS studies and guarantee the authenticity and confirmation of all activities (transactions) within the decentralized network.

\textit{C. Stringent Ownership Controls.} Control of genomic data should ideally be given to the owner of the data (the patient) or trusted third parties acting on behalf of the owner such as doctors. Necessary consent and access management mechanisms in the current centralized systems require more time and effort. A set of papers list the ability to control data as the main motivation for using blockchain. This is especially evident in genomic marketplaces that claim to allow individuals to control who has access to their data and for what purpose. As mentioned previously, there are also proposals (e.g.  \cite{mamo2020dwarna}) in which the patient consent is stored on the blockchain to empower patients and enforce their control over their own data. 

\textit{D. Immutability.} Immutable and tamper-proof data storage is the most desired property of blockchains in the selected papers. The immutability property in blockchain prevents the loss and alteration of data records which is essential in most genomic applications. The tamper-proof data structure of the blockchain, which relies on cryptographic hash pointers, prevents both accidental and intentional data tampering. Any changes to the confirmed blocks would make the blockchain inconsistent and can be discovered by any node in the network. This ensures a reliable and consistent shared ledger among untrusted or semi-trusted parties in the networks. Depending on the application requirements, on-chain storage can be leveraged to store data that needs to persist such as recording consent \cite{mamo2020dwarna} and providing an audit trail \cite{gursoy2020usingblockchain, ma2020efficient, pattengale2020decentralized, ozdayi2020leveraging}. However, data privacy must be carefully considered since the immutability property applies to on-chain data that is shared across all nodes and can be openly viewed if not encrypted. 

\textit{E. Smart contracts.} Smart contracts support various functions such as token generation and distribution, access control, and policy enforcement. Smart contracts are not part of all blockchain platforms, and not all applications require them. However, a percentage of papers have included smart contracts as a major part of their proposals. Commercial genomic data exchange platforms utilize smart contracts to generate the tokens that are used to incentivize individuals to share their genomic data. Both Encrypgen \cite{encrypgen} and DNAtix\cite{dnatix} use smart contracts to generate ERC-20 tokens that are used for payments. Smart contracts can also support the trading of genomic data for these generated tokens. In addition, enforcing an access policy by utilizing smart contracts is another motivation for employing blockchain. For example, \cite{carlini2020genesy} relies on smart contracts to allow access to the data and transfer the payment to the data owners. 

\textit{F. Reliability, availability, transparency and traceability.} A small percentage of papers specifically listed these blockchain properties as the main motivation for using blockchain. Reliability and availability are essential in certain applications such as online model learning in \cite{kuo2020ex}, which require data to be highly available to all nodes in the network. The importance of transparency and traceability are mostly highlighted in applications where the data owners are informed of how the data is accessed and by whom. These blockchain properties can be exploited to give data owners control and in turn gain their trust. For instance, \cite{mamo2020dwarna} emphasized the importance of transparency and argued that patients are more willing to contribute their genomic data for research purposes when they are informed about the use of their data.

\subsection{Current Blockchain-based Methods and Approaches in Genomics} \label{approaches}
There are various platforms \cite{kuo2019comparison}, storage systems \cite{benisi2020blockchain}, and privacy-preserving techniques \cite{zhang2019security} tailored to blockchain-based genomic data solutions. These technologies can be combined in different ways to deliver an application. In this section, we answer RQ4:What are the elements of blockchain technology used in genomic applications? What are the approaches or combinations of technologies used? \\
Table \ref{implementations} provides a summary of the methods used in the selected papers to implement a blockchain solution for genomics. The table is limited to the papers that provided an implementation or a proof-of-concept.
In the following subsections, we present a general architecture that covers most of the existing genomic blockchain systems, and then use this architecture to guide our presentation of the discussed work. 

\begin{table}
  \footnotesize
  \caption{Comparison of Blockchain implementations in genomics.}
  \label{implementations}
  \begin{tabular}{ m{1.2cm}  m{1.5cm} m{1.5cm} m{1.5cm} m{2.2cm} m{2.5cm}}
  \toprule
  \textbf{Paper} & \textbf{Blockchain Type} & \textbf{Platform} & \textbf{Storage method} & \textbf{Data type} & \textbf{Privacy} \\
  \midrule
  \cite{zhang2019enabling} & Private & Custom-made & Off-chain & Genotypes and phenotypes for GWAS & A gene fragmentation framework  \\
  \midrule
  \cite{ozercan2018realizing} & Private & Custom-made & On-chain \& Off-chain & Somatic cancer variation data & Hiding the germline
  variation \\
  \midrule
  \cite{Glicksberg2020} & Public & Ethereum & Off-chain (IPFS) & Clinical and genomic data & Anonymization  \\
  \midrule
  \cite{gursoy2020usingethereum, kuo2021benchmarking} & Private & Ethereum & On-chain & Pharmacogenomics & Restricted access to the private Proof-of-Authority blockchain\\
  \midrule
  \cite{mamo2019dwarna} & Private / permissioned & Hyperledger Fabric & On-chain \& Off-chain & Patient consent & User and study data stored off-chain with access control mechanism \\
  \midrule
  \cite{zhangy2018genie} & Public & Ethereum & On-chain \& Off-chain & Genetic and health data & Trusted execution environment (SGX) \\
  \midrule
  \cite{ileri2016coinami} & Public & Custom-made & On-chain \& Off-chain &  FASTQ files & Mixing reads from multiple genomes and inserting decoys \\
  \midrule
  \cite{gursoy2020storing} & Private & MultiChain & On-chain & Raw genomic data & Access control to private blockchain \\
  \midrule
  \cite{iyer2019ai} & Private / permissioned & Corda & On-chain \& Off-chain & Pharmacogenomics & N/A\\
  \midrule
  \cite{neto2020research} & Private / permissioned & BigchainDB & On-chain \& Off-chain & DNA sequence data & Anonymization \\
  \midrule
  \cite{kuo2020privacy,kuo2020ex, kuo2020anatomy, kuo2019fair, kuo2022detecting} & Private / permissioned & MultiChain & On-chain \& Off-chain & Aggregated data (the machine learning model) & Protected health information are stored off-chain\\
  \midrule
  \cite{pattengale2020decentralized, ozdayi2020leveraging, ma2020efficient, gursoy2020usingblockchain} & Private & MultiChain & On-chain & Access logs & N/A \\
  \midrule
  \cite{pachaury2021securing} & Private / permissioned & Hyperledger Fabric & On-chain & N/A & N/A \\
  \midrule
  \cite{warnat2021swarm} & Private / permissioned & custom-made & On-chain \& Off-chain & Aggregated data (the machine learning model) & privacy-preserving machine learning  \\
  \midrule
  \cite{ma2022phdmf} & Private / permissioned & Hyperledger Fabric & On-chain \& Off-chain & health \& genomic data  & on-chain data hashing \\
  \midrule
  \cite{miyachi2021hocbs} & Private & Ethereum & On-chain \& Off-chain & personal genomic data & off-chain data storage \& computation \\
  \midrule
  \cite{arroyo2021technological} & Private & Ethereum & On-chain \& Off-chain & personal genomic data & N/A \\
  \midrule
  \cite{park2021blockchain} & Private & Ethereum & On-chain \& Off-chain & personal genomic data & local differential privacy \\
  
  \bottomrule
  \end{tabular}
  \end{table}  

\subsubsection{General System Architecture}
\begin{figure}[h]
  \centering
  \includegraphics[width=10cm]{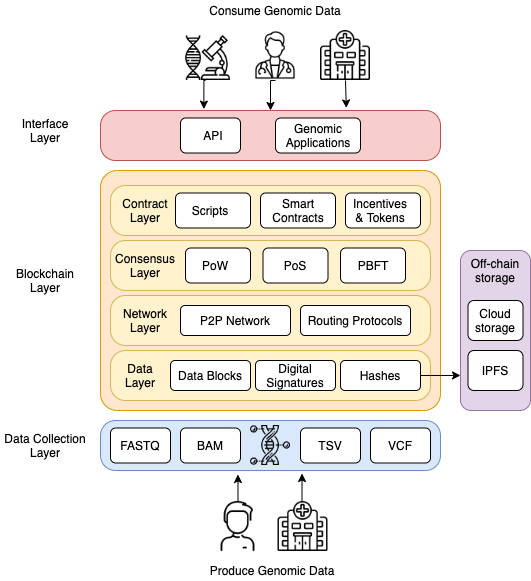}
  \caption{General architecture for blockchain-based genomic data sharing systems}
  \Description{}
  \label{fig:arch}
\end{figure}

Blockchain technology has been used for a variety of genomic applications as we have listed in the previous sections. Each of the selected studies has its own system architecture for the specific application at hand. However, most of the proposed solutions have architectural similarities that pave the way to generalization. In Figure~\ref{fig:arch}, we present a generalized architecture for systems that use blockchain for genomic applications. The design for this architecture is aimed at summarizing and covering a wide range of applications in genomics. The architecture consists of 6 layers: data collection, data storage, network, consensus, contract (decentralized application), and presentation. The layers in this architecture are comparable to the one/s in the existing blockchain literature but with some modifications to effectively illustrate the architectural components exploited in genomic applications.

In the first layer of the architecture, we assume that each node is responsible for collecting and sorting the genomic data which can come in different formats such as BAM or FASTQ. These nodes in the systems can represent an individual, researchers, or organizations that want to share genomic data. After the collection, the data is sent to the next layer for storage. In the storage layer, the data can be stored in different ways depending on the requirements, which is discussed in section \ref{storage_tech}. The data is then broadcasted to the network using a specified network protocol. According to our findings, the majority of the papers use a P2P network instead of the traditional client-server model. In the consensus layer, the nodes in the network come to an agreement on the state of the blockchain using a consensus protocol such as Proof-of-Work (PoW). The contract layer is where smart contracts are written and deployed to facilitate various application functions which serve as the backend of the application. The presentation layer is responsible for interacting with smart contracts and blockchain in general. 

\subsubsection{Blockchain Platforms}
A critical step in designing a genomic blockchain system is the selection of a suitable blockchain platform that would deliver the required functionality for the intended application. In this section, we present our analysis of the blockchain platforms used in the selected studies. Our results are only based on solutions that included a prototype, a proof of concept, or an implementation. We excluded studies that do not explicitly specify the blockchain platform and others that do not reveal their underlying platforms, especially in the context of commercial applications.

To identify the applicability or suitability of a blockchain platform, the requirements must be analyzed in the genomic data-sharing scenario. Based on the selected literature on genomic blockchains, the criteria used to identify the suitability of a blockchain platform are the following: 

\begin{itemize}
\color{blue}
  \item Public verifiability: is public verifiability required? Are the data contributors/users known in advance? The answer to these questions can influence the choice of blockchain platforms and whether it would be appropriate to use a public or private blockchain.
  \item The size and privacy of the data stored on-chain: is there a need to store large data on-chain? Is that data private or public? This can determine the type of consensus and block size that the platform provides. 
  \item Smart contracts: is there a need for smart contracts or on-chain computation? Since not all of the blockchain platforms support smart contracts, it is necessary to determine if on-chain computation or tokenization is required? 
\end{itemize}

\begin{figure}[h]
  \centering
  \includegraphics[width=10cm]{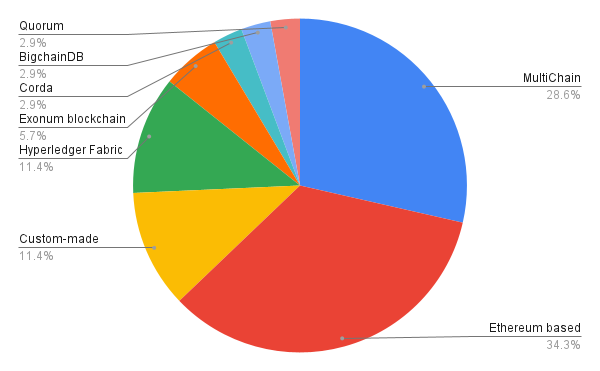}
  \caption{Blockchain platforms used in genomic data applications}
  \Description{}
  \label{fig:platforms}
\end{figure}

Figure \ref{fig:platforms} shows the blockchain platforms used in the selected studies. The majority of the papers proposed the use of either private or permissioned blockchains. Privacy, scalability, and cost are among the most cited reasons for this choice. The use of private blockchains lowers the risk of information leakage since data is only shared with a set of known semi-trusted individuals or institutions. In addition, private blockchains are more scalable and often use consensus mechanisms that do not require cryptocurrencies and transaction fees. The two main platforms used in genomic blockchain solutions are presented below.

\textit{Ethereum} \cite{ethereum} is a blockchain platform that facilitates building smart contracts and decentralized applications (DApps) that run on the blockchain network. Ethereum focuses on adaptability and flexibility, and, to achieve this, it supports Turing-complete programming language to build smart contracts easily. Although the Ethereum main blockchain network is public, in most of the selected papers, a private Ethereum blockchain is used. Similarly, Ethereum-based blockchain is employed in multiple genomic marketplaces such as \cite{zenome, Nebulagenomics, uribe2020privacy, dnatix}. In these use-cases, Ethereum smart contracts are used to facilitate access to genomic data files and the distribution of cryptocurrency. The previously mentioned CGT framework~\cite{Glicksberg2020} is another example solution that relies on Ethereum smart contracts for the distribution of genomic data. 

\textit{MultiChain}~\cite{multichain} is a blockchain platform to build and deploy private or permissioned blockchain networks. MultiChain focuses on providing features such as mining diversity, round-robin-based consensus, and data streams that allow on-chain data storage in a secure and efficient way. However, Multichain does not support smart contracts as do some other blockchain platforms. Multichain was used to build GeneChain \cite{encrypgen}. It was used in the IDASH competition~\cite{kuo2020idash} and thus all presented solutions~\cite{gursoy2020usingblockchain, ma2020efficient, pattengale2020decentralized, ozdayi2020leveraging} are based on MultiChain. The use of Multichain in these papers allowed efficient on-chain data storage of access logs. Additionally, MultiChain was used to implement ExplorerChain \cite{kuo2020ex, kuo2020anatomy} for the purpose of distributed and decentralized machine learning.

\subsubsection{Storage Techniques} \label{storage_tech}
There are two types of data storage that are compatible with blockchains, namely: on-chain and off-chain. The use of one or a combination of the two depends on the design requirements and consideration. This section describes the current blockchain storage approaches used in the literature to facilitate sharing and management of genomic data.

\textit{On-Chain Storage.} Storing data on-chain is achieved by simply adding the data (in binary format) to the transaction which effectively makes it part of the chain itself. This will eventually make the data immutable and highly available as the transaction will be distributed to all nodes in the network. However, some blockchains, especially public blockchains, have a strict limit on the size of each transaction making it difficult to store large data files. This is due to the fact that each full-node needs to have enough resources to store the ever-increasing amount of data being generated. On-chain data are also publicly accessible to all nodes in the network and therefore privacy of the stored data must be considered.

On-chain storage is most suitable for small data types that require immutable and tamper-proof storage. Data types that are commonly stored on-chain are meta-data and small genomic data. Small data types such as audit trail and observations of gene-drug interactions can be effectively stored on-chain as shown in \cite{gursoy2020usingethereum, gursoy2020usingblockchain, ma2020efficient, pattengale2020decentralized, ozdayi2020leveraging}. However, the majority of the papers only use on-chain storage for metadata \cite{Glicksberg2020, ozercan2018realizing, mamo2020dwarna, kuo2020anatomy, kuo2020ex, kuo2020privacy}.

On the other hand, whole-genome data which are stored in files such as BAM or VCF are large in size and difficult to store on-chain. There are attempts to store this data on-chain, such as genecoin \cite{genecoin}, and SAMchain \cite{gursoy2020storing}. Genecoin \cite{genecoin} proposed storing encrypted full-genome data on the bitcoin blockchain. While this approach is possible, it is not feasible due to the public nature of the bitcoin blockchain. Speed and scalability are severely affected by the need for each node to replicate these large data files. On the other hand,  SAMchain \cite{gursoy2020storing} uses a private blockchain (Multichain) to store and share sequence alignment maps on-chain using nested database indexing and compression techniques. Indeed, one of the main motivations of this paper is to prove that efficient storage and analysis are possible with a private blockchain. 

\textit{Off-Chain Storage.} The practical limitations of on-chain storage can be overcome by utilizing off-chain storage. In general, most off-chain storage techniques involve hashing (a piece of) data, which results in a small string that can be efficiently stored in the blockchain transactions or in a smart contract~\cite{benisi2020blockchain}. The actual data is then stored in either a centralized or decentralized storage system. The InterPlanetary File System (IPFS)~\cite{IPFS} is one of the most common approaches for decentralized off-chain storage. For instance, the CGT framework uses IPFS as off-chain storage. The data is stored on the IPFS servers and only a strong hash (SHA-256) is added to the blockchain. The hash uniquely defines the entire state of all data submitted from the steward at that exact point in time. Similar hashing of raw data is used in \cite{zhangy2018genie} to validate that off-chain data has not been tampered with. Central off-chain storage system was proposed in Genesy~\cite{carlini2020genesy} and CrypDist~\cite{ozercan2018realizing}. Cloud storage is used to store the data and only the hash of the data is stored on-chain. A transaction is created with every file containing a hash pointer to the data on the cloud. Large BAM files are stored off-chain in this scenario, while the metadata about the patient and the access rules, such as phenotypic and environmental information, is stored on-chain.

\subsubsection{Security and Privacy Protection}
\label{subsec:secpri}
In genomic applications, the privacy requirements vary depending on the data type. Somatic variants are generally considered to be non-sensitive and do not require any privacy protection. On the other hand, germ-line variants are private and therefore, privacy protection is essential and it is enforced by existing regulations such as HIPAA and GDPR. In addition, genomic and medical information (extracted from EHR) are often combined for genotype-phenotype analysis. This creates another point of privacy concern because when multiple data points are combined, there is a risk of correlation attacks through which the identities of patients can be revealed by combining multiple identifying data points \cite{naveed2015privacy}. With these privacy concerns, it is necessary to look closely at how this challenge is currently tackled in genomic blockchains. 

The majority of the papers rely on the security of the cryptographic protocols employed in the basic blockchain implementation which include hash pointers, Merkle trees, digital signatures, public key infrastructures (PKI), and consensus protocols \cite{zhang2019security}. As previously discussed in Section~\ref{block_background}, the combination of these techniques provides a robust decentralized system that can withstand malicious tampering.

Privacy through data anonymization is one of the approaches used in most papers. CrypDist \cite{ozercan2018realizing} and CGT \cite{Glicksberg2020} achieve privacy by sharing only somatic variants and removing the personally identifiable patient data and the private germ-line variants. While anonymization provides a certain level of privacy, it does not guarantee protection against future re-identification attacks \cite{rocher2019estimating}. Another approach to privacy is the use of private blockchains. Gursoy, et al \cite{gursoy2020storing} uses a private blockchain that requires permission to access the data within the blockchain. With controlled access, there is a limited number of security issues.

In the machine learning use cases such as \cite{warnat2021swarm, kuo2020ex, kuo2020anatomy, kuo2019fair}, privacy is addressed by training the model locally and then merging the model parameters. The authors use blockchain as a way to coordinate the process of distributing the model instead of a central server that could potentially breach the confidentiality of the data. However, the authors point out that risks of re-identification still exist and further privacy-preserving techniques such as differential privacy are required for optimal privacy protection.

Genie \cite{zhangy2018genie} presents a blockchain-based solution to AI model training with the added security of a trusted execution environment namely, Intel Software Guarded eXtensions (SGX). The secure enclave is used to train the models and therefore privacy is preserved by protecting the raw data while still allowing the sharing of insights from it. These security and privacy protection techniques are combined with the transparency, control, and verifiability of blockchain in the proposed solution.

Zhang et al \cite{zhang2019enabling} provided a different approach to privacy. In the proposed framework, the gene sequence of one individual is fragmented into $n$ pieces and distributed to analysis nodes, which run a specified analysis on the given part of the data. The fragmentation lowers the probability of re-identification in each analysis node and makes sure each fragment of the data is unidentifiable.

\cite{grishin2019citizen} uses multiple cryptographic techniques that are added to the data-sharing system. Homomorphic encryption is used to encrypt the data and process it. Differential privacy is also used to add another layer of privacy and prevent the re-identification of individuals. This is done by adding noise to obfuscate the query results.  

\subsection{Open Issues and Challenges}\label{sub:openissues}
The following is a list of current challenges and limitations in applying blockchain in genomics which answers the predefined RQ5: What are the challenges and limitations when applying blockchain in genomics? Have these limitations been addressed by the authors? What has been specified for future research? 

\subsubsection{Adoption barriers} 
Public adoption of blockchain in genomics and healthcare, in general, requires ease of use and software stability. Blockchain platforms are continuously changing and require expert knowledge for adoption. This has been noted as one of the major challenges in implementing solutions in most papers. Instability and lack of user-friendly interfaces are major barriers to public adoption and therefore limit its use to tech-savvy individuals. 

\subsubsection{Interoperability} 
For the adoption of blockchain in genomics, organizations need to integrate and connect blockchain platforms with existing non-blockchain platforms. This problem is aggravated as there is a wide range of blockchain implementations that are not necessarily compatible with each other. Interoperability between blockchains reduces the dependence on a single blockchain platform. Most of the presented solutions in genomics rely on a specific blockchain platform and its features. A multi-blockchain approach, which doesn't rely on a specific blockchain platform, would provide better scalability and remove the security risks associated with the used platform. However, this approach is currently complicated and involves complex cross-chain communication. Research in the area of blockchain interoperability is growing and multi-blockchain approaches might be feasible in the future \cite{belchior2020survey}. 

\subsubsection{Smart Contract Security}
Blockchains that support smart contracts enable building rich applications that are not limited to financial transactions. However, the increased functionality provided by smart contracts exposes the system to more possible attacks such as the DAO attack \cite{mehar2019understanding} in 2016. The number of discovered smart contract vulnerabilities is increasing \cite{chen2020survey}, and they can be costly, either in terms of financial loss or data privacy loss. Smart contracts were used in a number of papers to achieve various important functionality, such as granting access to private data. One of the challenges in deploying such smart contracts to handle actual patient data is the security risk associated with them. Following best practices and performing security audits might reduce this risk, and research in smart contract security is still ongoing \cite{chen2020survey}.

\subsubsection{Data Privacy} 
Even though privacy in blockchain has been studied extensively in the literature, privacy issues with blockchain in genomic applications have not been fully addressed. There is a need to examine some areas of privacy, especially with the anonymity of users and re-identification through correlation attacks. While re-identification is sometimes required in research settings, it is essential to prevent the disclosure of patient data for any other purpose. For instance, re-identification is required when additional materials are needed for further research into the case. However, re-identification to reveal patients' private data should be prevented. Identity and transaction privacy is another challenge. Ideally, identifying a user based on specific interactions with organizations should not be possible. On-going research in using zero-knowledge proofs (ZKPs) \cite{partala2020non} have demonstrated the possibility of achieving this in financial applications. Moreover, risks of re-identification associated with open data sharing are still present even after the full de-identification process as has been shown in \cite{rocher2019estimating}. There is still a risk that more advanced re-identification attacks might emerge in the future. Therefore, it is an open question whether other privacy-preserving mechanisms can be applied to ensure privacy against future attacks.

\subsubsection{Reproducibility and validity} 
Replicating the study to determine its validity is essential in genomic and research in general. Researchers or auditors would want to find exactly the same data without change after a number of years to replicate the study. In cases where blockchain facilitates data sharing for scientific studies, it is important to address this particularly when off-chain storage is used. Aside from using decentralized off-chain storage such as IPFS, this problem manifests especially in solutions that use centralized off-chain storage. For instance, the data might be removed from the cloud provider in the future. The data should follow the FAIR principles, however, there seems to be a lack of focus on this aspect in genomic blockchains. 

In addition, data redundancy might occur where patients exist in multiple organizations with different assigned IDs. A challenge with decentralized sharing/analysis of genomic data is the possibility of publishing duplicate data, where the same patient data is shared but with different anonymous identities. This is a problem especially if this data is intended for research purposes and it can affect the validity of the study. The Cancer Gene Trust \cite{Glicksberg2020} highlighted this limitation and tried to eliminate this problem with a rule-based scoring system and 2 reviewers. There are existing techniques to identify duplicate records in different databases, such as \cite{laud2018privacy} which performs privacy-preserving record linkage on several databases using secure multiparty computation. However, further research is still needed to address this problem in a decentralized blockchain network. 

\subsubsection{Verifiability}
One of the issues associated with distributing processing tasks to untrusted parties is verifying the accuracy of the results. One possible solution is outsourcing the same analysis task to multiple analysis nodes and then the results can be compared to ensure correctness. However, the cost of this approach can be high, especially if the same task is distributed to a large number of analysis nodes. Therefore, a practical and scalable verification method to ensure that outsourced computation/analysis is indeed correctly computed by untrusted nodes in a blockchain network is still an open issue that requires further examination. There are significant advancements in the field of verifiable computation \cite{yu2017survey} which can be explored in a blockchain setting.

\subsubsection{Key management} 
It is challenging to ensure that the patients (data owners) are able to manage securely their keys and identity, especially with data related to individual health. Once patients have full control over their data, education mechanisms must be put in place for the patients, in order to provide them with valuable insights regarding best data management practices. Moreover, proper key management schemes need to be put in place along with mechanisms for “break glass” access to genomic/healthcare data in emergency settings.

\subsubsection{Ethical Challenges} 
The rise of genomic marketplaces raises some ethical concerns as discussed by Ahmed et al \cite{ahmed2019dna} and Defrancesco et al \cite{defrancesco2019your}. These authors argue that informed consent is questionable when a monetary incentive is involved and it can lead to mindless data sharing. It is yet to be known if these financial incentives would actually work in attracting more people to share their private data for research purposes, but perhaps alternative non-monetary incentives should be explored. For Instance, Mofokeng et al \cite{mofokeng2018future} showed how digital collectibles in the form of non-fungible tokens can be used to incentivize citizens to participate in wildlife conservation. Therefore, further research into non-monetary incentives that encourage participation in genomic research for the purpose of advancing medical research might prove to be effective. 

\section{Discussion} \label{discussion}
In this section, we discuss the major findings of this review, the challenges, limitations, and propose several future research directions in applying blockchain in genomics.

\subsection{Principal Findings}
Our results suggest that there is an increasing interest in applying blockchain in genomics since the number of publications has increased rapidly each year since the inception of blockchain and the developments in genomics. The use of blockchain technology in genomic applications is becoming common in both commercial and non-commercial settings. Commercial applications focus on the need for user control (i.e. data ownership) and at the same time enable data owners to profit from its use and sharing. Rewarding a form of cryptocurrency to the data owners is commonly used as an incentive mechanism to attract more contributions. Non-commercial applications provide solutions for sharing, processing/analysis, secure storage, and access control to genomic data. These solutions aim to facilitate easy, efficient, multi-organizational genomic data sharing for the purpose of advancing genomic research. The main motivations for using blockchain that were highlighted in the papers are \emph{immutability}, \emph{decentralization}, and \emph{access/usage control}. 

Our findings indicate that private or permissioned blockchains are the most common blockchain types, and Multichain and Ethereum are the most common platforms used in genomics. The storage techniques used vary depending on the requirements. \emph{On-chain storage} is mainly used for small data types or hashes/pointers of the data that are stored off-chain. \emph{Off-chain storage} is often used for large data files or for data that requires strict access control. In these cases, either cloud storage or other decentralized file systems such as IPFS is used. Further investigation of \emph{data compression} techniques suitable for decentralized storage, sharing, and processing of genomic data is needed.

Since genomic data is long-lived (i.e. valid for a very long time), data security and privacy are among the principal concerns. Most of the existing papers utilize existing protection mechanisms that are part of the blockchain itself such as consensus mechanisms and digital signatures for the purposes of integrity, confidentiality, and availability. Privacy is protected by either anonymization or the use of private blockchains with controlled access. There are, however, solutions that proposed adding cryptographic techniques to further protect the privacy of the data.

\subsection{Review Limitations}
While this review aims to provide a comprehensive overview of the current state of the art, limitations still exist. The main focus of this review is to cover blockchain applications in genomics, therefore, we do not cover other healthcare-related studies in which blockchain is proposed as a solution. Our search strategy was aimed to capture only papers that specifically emphasize genomic data applications. We also focused on the most popular commercial genomic platforms, and therefore we only reviewed the first 200 results returned from the initial Google search. The documentation in white papers is somewhat limited and could change as the technology matures. We observed that details in some white papers change over time. We tried to overcome this by including the details of the latest versions of these white papers. Additionally, we have not run any testing on the proposed solutions to verify the claims stated in the papers. 

\subsection{Future Directions} \label{future_directions}
To achieve large-scale deployments and adoption of blockchain in genomics, we point out that further exploration is needed for this technology to mature. There are multiple challenges and interesting problems in blockchain research that need to be addressed. In this section, we highlight some possible future research directions that we observed after conducting the review.

Based on our results, the use of cryptographic protocols for privacy-preserving analysis and computation is limited in current blockchain-based approaches in genomics. Cryptographic protocols such as Multi-Party Computation (MPC) can enhance privacy and when combined with blockchain and smart contracts, it can help address the security, trust, and verifiability issues in distributed analytics \cite{zhong2019secure}. Current research in this field has shown the feasibility of performing distributed privacy-preserving analysis on blockchain \cite{ghadamyari2019privacy,zhou2021using}. Attribute-based encryption (ABE) is yet another powerful technique that can be leveraged to enhance the security and privacy in blockchain solutions \cite{sun2018decentralizing}.

Furthermore, we observed that the solutions are all based on a limited number of blockchain platforms. While these platforms are popular and have proved to be effective, perhaps experimenting with other emerging blockchain platforms might show some interesting results. Another point to consider is that all papers included in this review propose the use of blockchain and not DLT. The use of DLT has been proposed for other applications such as IoT, but it seems to be missing in the genomic literature. Further research into the use of DLT is required to assess the feasibility of this technology in genomics. 

Recent development in zero-knowledge (ZKP) based off-chain computations is promising and would open the door for more applications to be built on the blockchain. In short, off-chain computation is an alternative state transition model where the function is computed off-chain by one of the nodes and then the resulting state persists on-chain after verifying the computation (verifying the ZKP proof). Arbitrum is an alternative solution which uses an incentive-based verification of off-chain computations \cite{kalodner2018arbitrum}. Given that genomic data is rather large in size and heavy computation is often required for genome analysis tasks, we believe there is an opportunity for further experimentation with these off-chain computations methods in genomics.

We also see a lack of focus on the aspect of trust and the emerging technologies that can be used to build it such as Decentralized Identifiers (DIDs) \cite{reed2020decentralized} and Verifiable Credentials (VCs) \cite{sporny2019verifiable}. Utilizing blockchain to build a decentralized trust infrastructure, which can help in addressing other challenges such as data integrity and privacy. Following the trust-over-ip principles \cite{davie2019trust} in the genomic applications can provide a layer of trust when sharing data for research purposes.

The use of blockchain in genomics is still in its early stages and there are many other use cases to be explored. We expect that the development of blockchain-based solutions would change the current genomic ecosystem. As one of the aims of using blockchain is to empower patients to control their data, their role in data sharing will be significant. Increased trust and automated processes provided by blockchain and smart contracts would scale the amount of data being shared. In addition, blockchain allows the design of incentives that can facilitate sharing, storing, and processing genomic data in a fair way and for the ultimate purpose of advancing our knowledge of the human genome. 

\section{Conclusion}
There is an increasing number of papers proposing blockchain-based solutions to enable the storage, sharing, and processing of genomic data. A similar trend can be observed from the large number of commercial/non-commercial blockchain applications that aim to enable genomic data exchange. In this paper, we provided a comprehensive overview of the existing efforts in this area.

Our study employed a taxonomy in which genomic applications of blockchain are classified into commercial, and non-commercial applications. Non-commercial applications have been further categorized according to their specific goals, namely data sharing, analysis, secure storage, and access control. After providing certain details about each application, we described the advantages and drawbacks of the proposed approach and provided a comparison between the proposals concerning the blockchain platform selection, how data is stored, shared, and protected in the proposal. Software instability, interoperability, and the security risks associated with the rigidity of smart contracts are some of the highlighted challenges. Another challenge is protecting the privacy of the data and the identities of the data owner. Privacy-enhancing technologies such as those mentioned earlier can be beneficial when applied in these settings.

Our results suggest that immutability and decentralization are the main motivations for blockchain use in this context. We observed that empowering data owners to control their data is a common argument in the papers. In most papers, blockchain is used to give control of the data to organizations or individuals (patients). 

We also observed that the applications or use-cases of blockchain in genomics are rather limited compared to financial applications although there is a huge potential in exploring alternative use-cases. We identified a list of open issues/challenges from the perspective of the existing proposals (Section~\ref{sub:openissues}). We recommend experimenting with blockchain-based distributed analytics (i.e. processing) in genomics. In addition, evaluating the performance of various privacy-enhancing technologies such as homomorphic encryption, multi-party computation, zero-knowledge proofs, and off-chain computation can shed some light on the feasibility of privacy-preserving distributed analytics in blockchain networks. Another promising future direction is exploring blockchain-based trusted and verifiable access to genomic data. The trust-over-ip principles \cite{davie2019trust} can be utilized to create and manage decentralized identities for researchers. 

Our general impression is that genomic applications on blockchain are still in their early stages of research and development, and require further transformation both socially and technologically in order to be adopted. Recent efforts to promote the use of blockchain in genomics such as \cite{IDASH} can help accelerate the adoption of this technology by showcasing the potential and feasibility of using it in various genomic applications.

\medskip

\bibliographystyle{ACM-Reference-Format}
\bibliography{main}

\end{document}